# Diffusion-Weighted MR imaging:
# Clinical applications of kurtosis analysis to prostate cancer

Andrea Barucci[1,*], Roberto Carpi[2], Marco Esposito[2],
Maristella Olmastroni[2], Giovanna Zatelli[2]

[1] Istituto di Fisica Applicata "Nello Carrara" del CNR (IFAC-CNR)
[2] Azienda USL Toscana Centro, Piazza Santa Maria Nuova 1, Firenze, Italy

[*] A.Barucci@ifac.cnr.it

## 1 - Introduction

Magnetic resonance imaging technique known as DWI (diffusion-weighted imaging) allows measurement of water diffusivity on a pixel basis for evaluating pathology throughout the body and is now routinely incorporated into many body MRI protocols, mainly in oncology [1-5]. Indeed water molecules motion reflects the interactions with other molecules, membranes, cells, and in general the interactions with the environment. Microstructural changes as e.g. cellular organization and/or integrity then affect the motion of water molecules, and consequently alter the water diffusion properties measured by DWI. Then DWI technique can be used to extract information about tissue organization at the cellular level indirectly from water motion.

In general the signal intensity in DWI can be quantified by using a parameter known as ADC (Apparent Diffusion Coefficient) emphasizing that it is not the real diffusion coefficient, which is a measure of the average water molecular motion. In the simplest models, the distribution of a water molecule diffusing in a certain period of time is considered to have a Gaussian form with its width proportional to the ADC [6,7]. However, water in biological structures often displays non-Gaussian diffusion behavior, consequently the DWI signal shows a more complex behavior that need to be modeled following different approaches.

In this work we explore the possibility to quantify the degree to which water diffusion in biologic tissues is non-Gaussian introducing the AKC parameter (Apparent Kurtosis Coefficient). DKI was first described by studies in 2004 [8] and 2005 [9] and initially was applied exclusively for brain imaging [10-12], while in recent years some studies have shown the feasibility of applying DKI at multiple extra-cranial sites [13-18].

In this work we have realized DWI non-Gaussian diffusion maps to be used in the clinical routine along with standard ADC maps, giving to the radiologist another tool to explore how much structure inside a voxel is organized.

In particular in this work some prostate DWI examples have been analyzed and will be shown. References to other studies using DKI in detection and characterization of prostate cancer can be found here [1,14,19–38,49,61-63].

## 2 - An introduction to Water Diffusion

A complete description of the diffusion theory and DWI technique is beyond the scope of this article, so here we introduce some important concepts and equations, leaving some references [2-4] for the interested readers.

Diffusion measurements in MRI usually can be performed using the standard diffusion-weighted pulse sequence (spin-echo echo-planar imaging) [39-41], obtaining images called DWI. DWI is performed by serially imaging the same tissue while varying the degree of water diffusion sensitization. The imaging gradient strength, direction, and temporal profile affect sensitivity to diffusion and are commonly reduced to a single simplified parameter referred to as the b-value [unit: s/mm$^2$]. The images obtained at different b-values are subsequently used for computing a parametric map that allows quantitative assessment of the tissue's water diffusion behavior.

In this context the corresponding echo attenuation in a voxel can be expressed as

$$S(b) = S_0 \times \exp(-b \times ADC) , \tag{1}$$

where S is the signal intensity (a.u.), depending upon the apparent diffusion coefficient (ADC) and the diffusion-sensitizing factor, which can be calculated for a spin echo sequence with rectangular diffusion-encoding gradients as follows [40]:

$$b = \gamma^2 G^2 \delta^2 \left(\Delta - \frac{\delta}{3}\right). \tag{2}$$

Here, $\delta$ is the duration of one diffusion-encoding gradient lobe, $\Delta$ is the time interval between the leading edges of the gradient lobes, $G$ is the strength of the gradient, and $\gamma$ the gyromagnetic ratio.

Then a fit on a voxel basis of equation (1) as a function of different b-values gives the ADC map that can superimposed on the standard anatomical images in order to obtain more information on the tissue under investigation.

However, biological tissues are highly heterogeneous media that consist of various compartments and barriers with different diffusivities. In terms of its cytohistologic architecture, a tissue can be regarded as a porous structure made up of a set of more or less connected compartments in a networklike arrangement.

The movement of water molecules during diffusion-driven random displacement is then impeded by compartmental boundaries and other molecular obstacles in such a way that the actual diffusion distance is reduced, compared with that expected in unrestricted diffusion. This is the reason for which the classical model of diffusion used in MRI is not always correct and must be thought as an approximation in many situations. Instead water in biological structures shows often non-Gaussian diffusion behavior. As a result, the MR signal intensity decay in tissue is not a simple mono-exponential function of the b-value [1,15,42] as described in equation (1).

Several approaches have been used to model the nonlinear decay of DWI signal intensity when more than 2 b-values are acquired. These approaches include bi-exponential fitting, from which 2 components that hypothetically reflect 2 separate biophysical compartments can be derived [43], stretched-exponential fitting, which describes diffusion-related signal intensity decay as a continuous distribution of sources decaying at different rates [44], and diffusional kurtosis analysis, which takes into account non-Gaussian properties of water diffusion by measuring the kurtosis [9].

Kurtosis represents the extent to which the diffusion pattern of the water molecules deviates from a perfect Gaussian curve. Unlike the bi-exponential model, the stretched-exponential and the kurtosis methods do not make assumptions regarding the number of biophysical compartments or even the existence of multiple compartments [45]. From the kurtosis analysis the apparent diffusion coefficient (ADC) and the apparent kurtosis coefficient (AKC) can be estimated, which are phenomenological parameters [1] supported by observations and with no direct biophysical correlation.

What we can observe is that more organized is a structure of parenchyma, much more constrains a water molecules can explore during diffusion process [1,42,46-48]. Furthermore any modification of cellular arrangements, cell size distributions, cellular density, extracellular space viscosity, glandular structures, and integrity of membranes or to measure any modification of macromolecule's concentration, translates in some features of ADC and AKC. However an interpretation of ADC and AKC is still now no straightforward [9,50-58].

The AKC parameter (adimensional) can be inserted in the mathematical signal formulation as follows:

$$S(b) = S_0 \times \exp\left(-b \times ADC + \frac{1}{6} AKC \times b^2 \times ADC^2\right). \tag{3}$$

This quadratic model shows a better agreement in many tissues as shown in the example of Fig. 1.

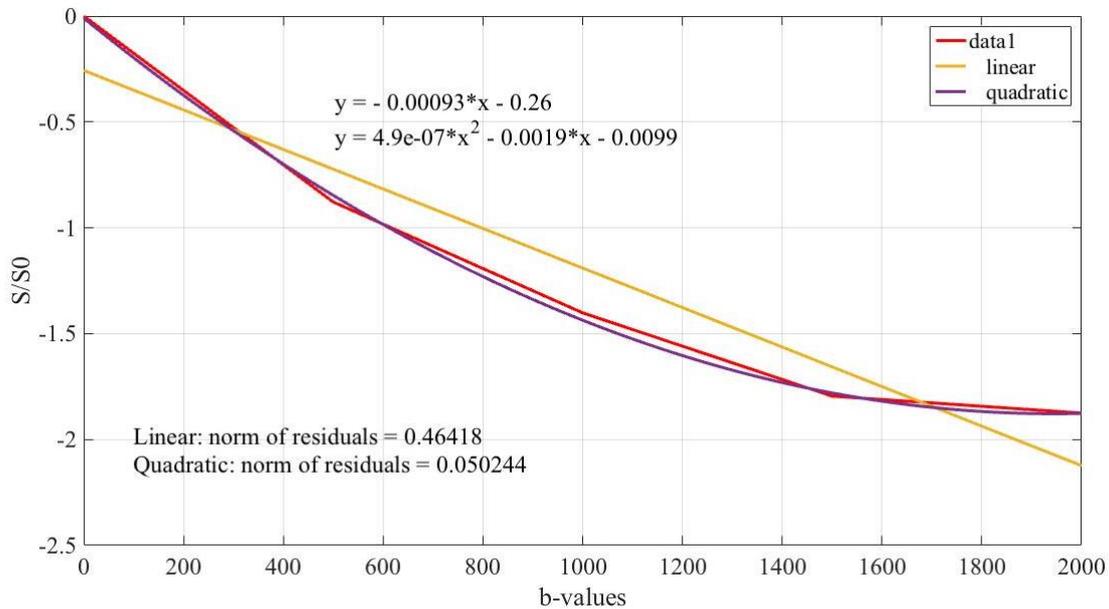

**Fig. 1** - Examples of data fitting with linear and quadratic models. The quadratic model incorporating the Kurtosis term shows a better norm of residuals in respect to the linear model. Data coming from a patient affected by prostate cancer.

AKC equals 0 when water is experiencing completely Gaussian diffusion [1], while biologic tissues tend to exhibit AKC values between 0 and 1. Studies also suggest lowering of K in the setting of post-treatment tumor necrosis [59,60]. Post-processing software commonly applies a maximal possible upper limit for AKC, above which the value is likely to represent an outlier due to motion, noise, or other artifact [1,9,55].

## 3 - Materials and Methods

Using the Philips Achieva 1.5 T available for clinical routine use at the Santa Maria Nuova Hospital in Florence we acquired a dataset of 20 patients affected by suspected prostate cancer calculating for each the ADC and AKC maps. A set of 5 b-values (0, 500, 1000, 1500, 2000 s/mm$^2$) was chosen as a trade-off for clinical use and best signal-to-noise ratio in DWI [1]. Usually b-values above 1000 s/mm$^2$ are necessary to successful capture the non-gaussian behavior.

Special software has been developed in the MATLAB framework in order to open and elaborate the DICOM images coming from the MR scanner. This software allows the data elaboration of DWI images realizing ADC and AKC maps (Figs. 2-6), at the same time introducing some post-processing tools (as moving average filter or interpolating algorithm) in order to support radiologists in the images interpretation.

## 4 - Results

DWI images have been acquired for all the patients' dataset, estimating ADC and AKC on voxel basis using Eq. (2).

Radiologists on clinical practice, cross-correlating these results with the other coming from standard MRI examination and patient clinical report, have used the obtained ADC and AKC maps.

An example of ADC and AKC maps is shown in Fig. 1.

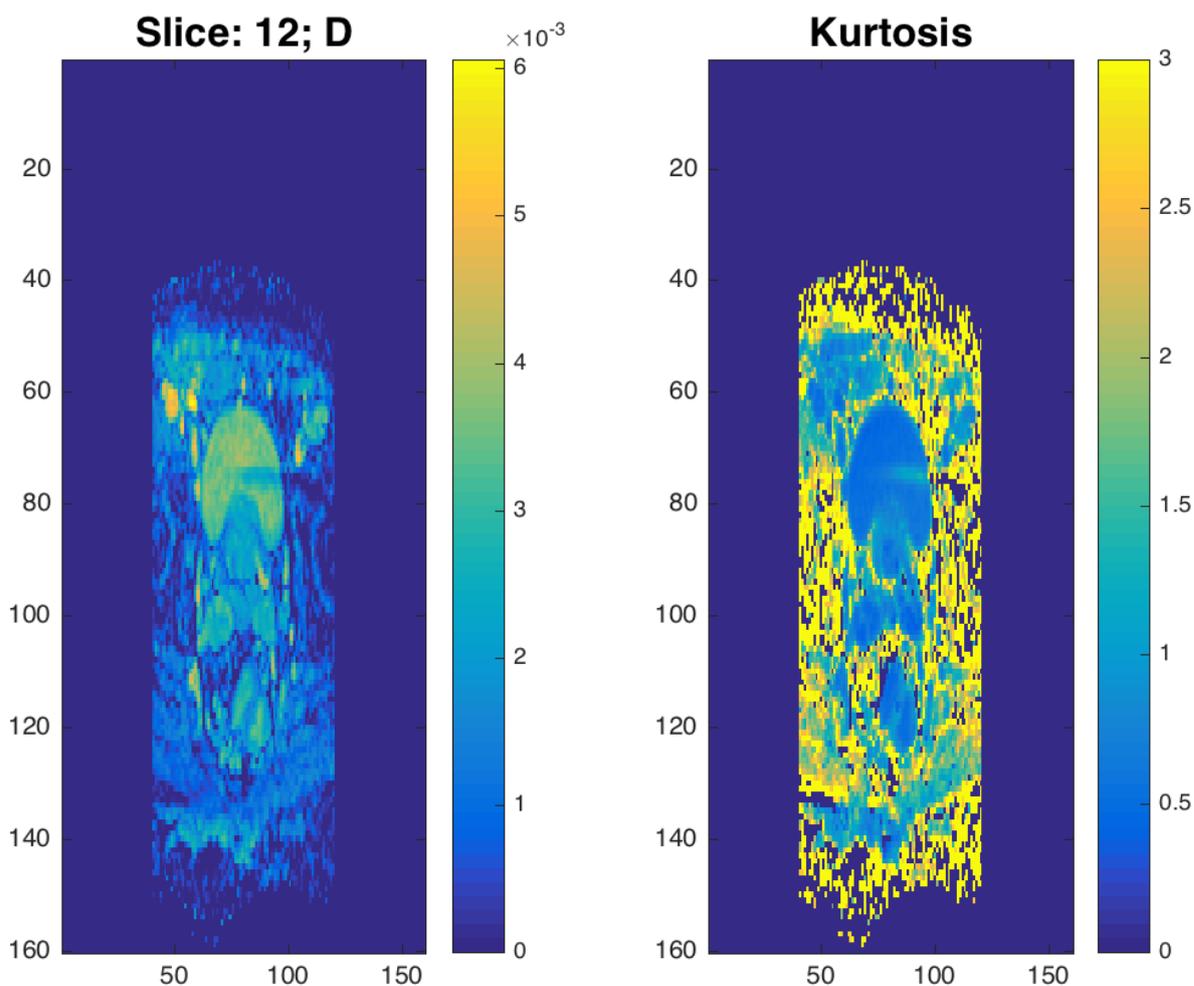

**Fig. 2 -** Example of ADC ("D") and AKC ("Kurtosis") for a patient slice.
In this case the color is different from the standard gray scalar of radiology.

Figure 3 is an example of ADC and AKC maps before and after post-processing with data interpolation. The color scale in this case is the standard for radiologists. This is just an example of the software developed for DWI data analysis.

In Fig. 4 and Fig. 5 two examples of ADC tridimensional view are shown for two patients, while for Patient 2 the AKC tridimensional view is shown in Fig. 6.

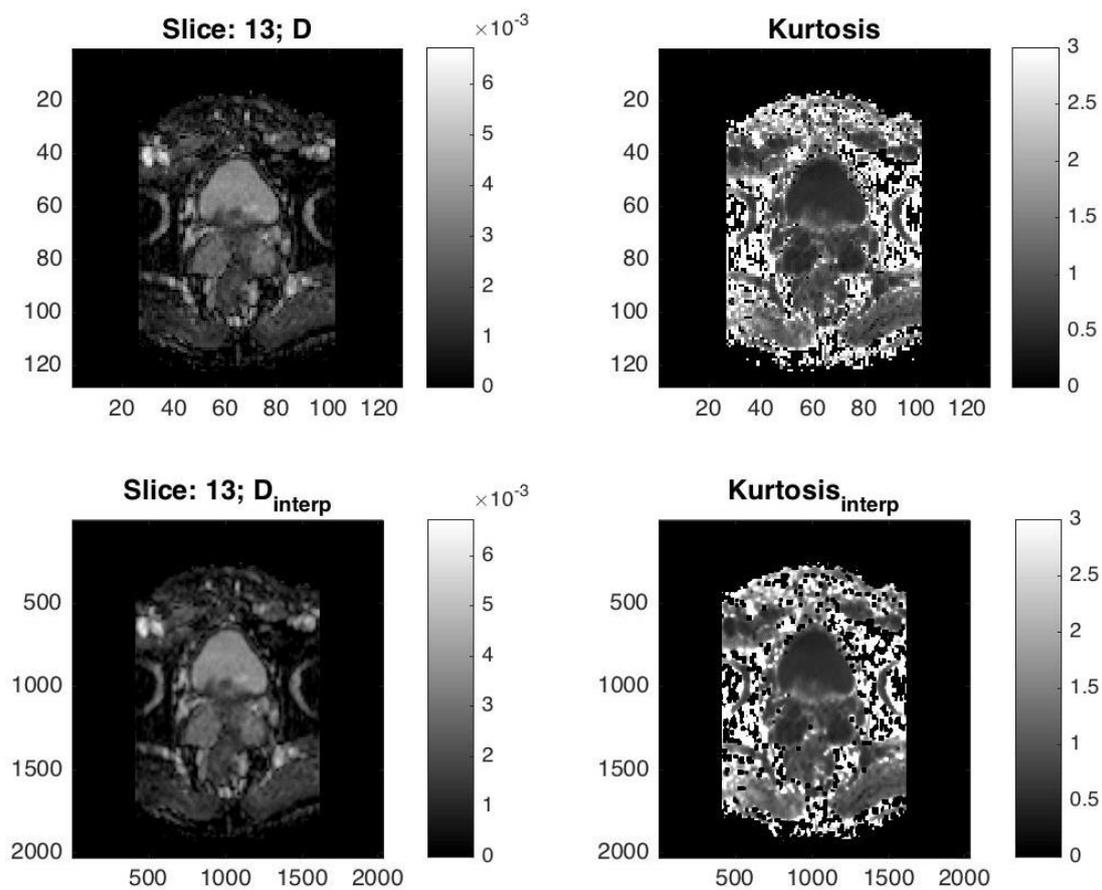

**Fig. 3 -** Examples of ADC ("D") and AKC ("Kutosis") parameters for a patient slice. In the first line the original coefficients, while in the second line the maps were interpolated in order to obtain a better resolution.

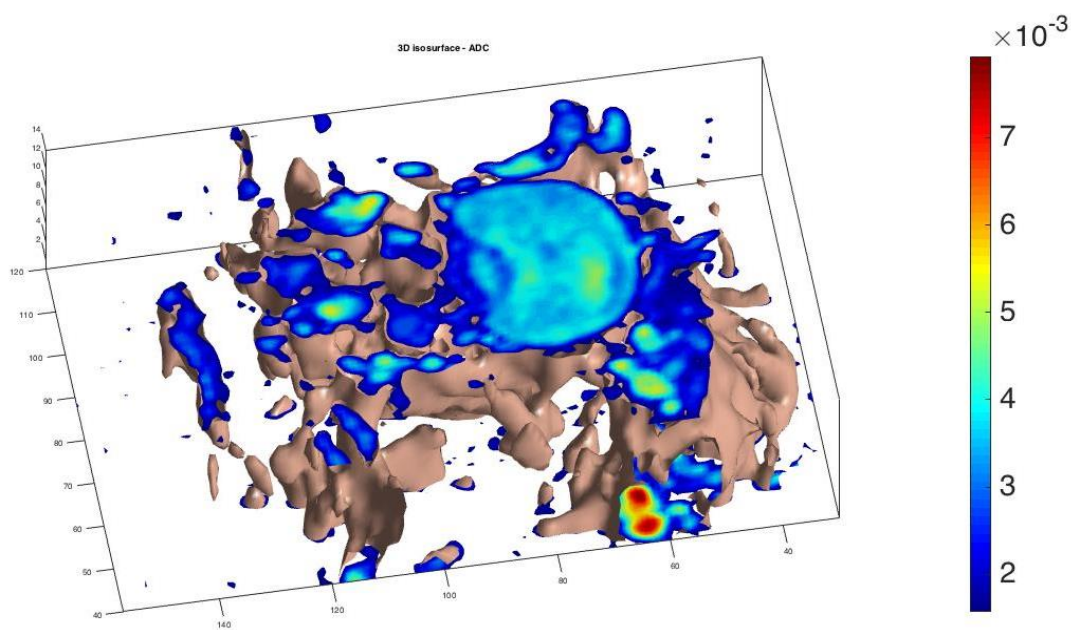

**Fig. 4 -** Example of a tridimensional ADC map for patient 1.

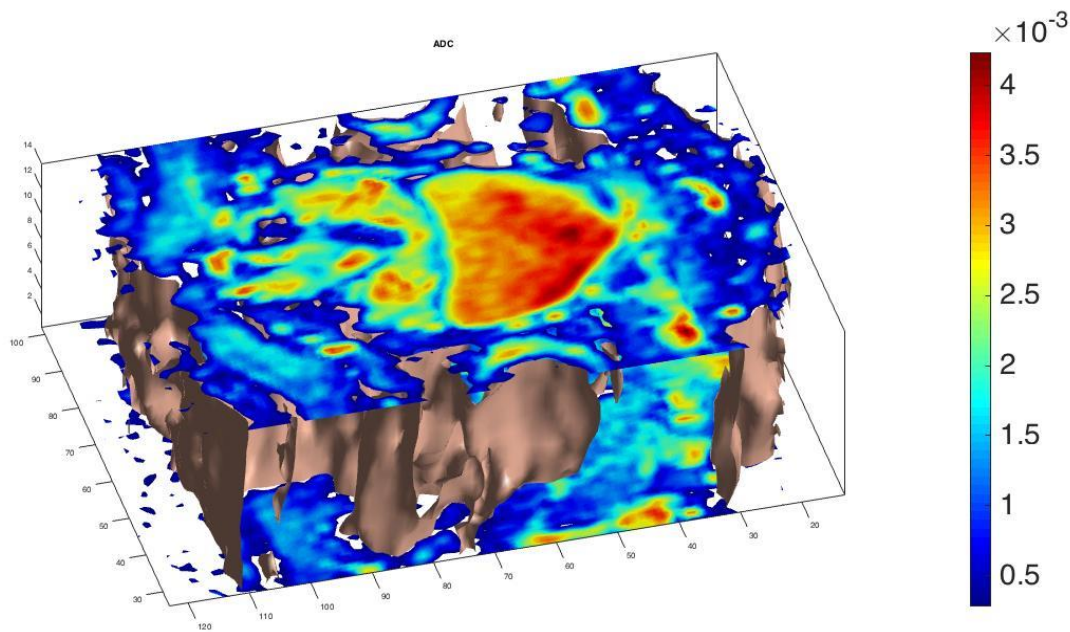

**Fig. 5 -** Example of ADC for patient 2 in a tridimensional view.

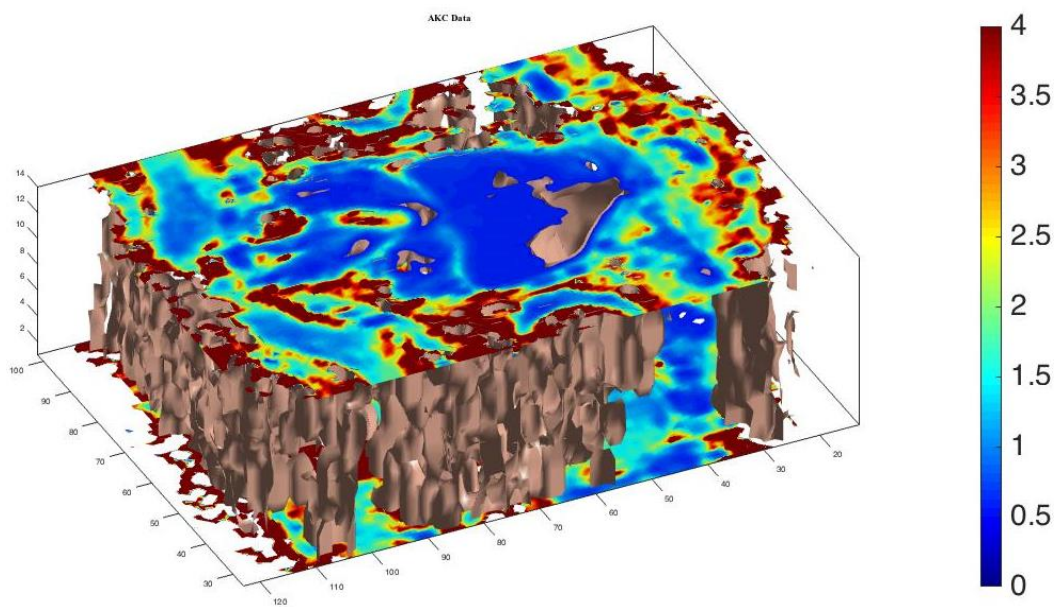

**Fig. 6 -** Example of AKC tridimensional view for patient 2.

## 5 - Conclusions

DWI non-Gaussian analysis has shown the potential to become a powerful tool in supporting radiologists in the clinical practice.

However much work remains to be done to fully understand the mechanisms underlying non-Gaussian diffusion, and the precise bio-structural significance of AKC in relation to microstructural properties of tissues.

In this framework we are working on a new and different approach based on the theoretical physics of diffusion in complex medium [64-67]

At the same time we are working on some kind of nanoparticles as a new theranostic agent for MRI applications, in particular trying to understand if nanoparticles can be revealed by diffusion-MRI techniques, looking at the change in water motion due to the presence of nanoparticles in the environment [68,69].

## 6 - Acknowledgments

The authors wish to thank Fulvio Ratto, Sonia Centi, Francesco Baldini, Ambra Giannetti and Roberto Pini, from IFAC-CNR for supporting and fruitful discussions, and the project IRINA - "Imaging Molecolare di risonanza magnetica della biodistribuzione di nanoparticelle e vettori cellulari per applicazioni teranostiche" by "Ente Cassa di Risparmio di Firenze" for financial support [Rif. Pratica n. 2015.0926, Sede Legale: via Bufalini 6, 50122 Firenze, www.entecarifirenze.it].